\documentclass[useAMS,usenatbib,usegraphicx]{mn2e}

\usepackage{xcolor}

\usepackage{bm}
\usepackage{amsmath}
\usepackage{amssymb}
\usepackage{xspace}
\usepackage{lineno}
\usepackage{url}
\newcommand{\maglites}{{MagLiteS}\xspace}

\newcommand{\code}[1]{\texttt{#1}\xspace}

\title[Discovery of Carina~II and III]{Discovery of two neighboring
  satellites in the Carina~constellation with MagLiteS}

\author[G.~Torrealba et al.]{
\parbox{\textwidth}{
\Large
G.~Torrealba$^{1,2}$,
V.~Belokurov$^{1,3}$,
S.~E.~Koposov$^{4,1}$,
K.~Bechtol$^{5}$,
A.~Drlica-Wagner$^{6}$,
K.~A.~G.~Olsen$^{7}$,
A.~K.~Vivas$^{8}$,
B.~Yanny$^{9}$,
P.~Jethwa$^{1}$,
A.~R.~Walker$^{8}$,
T.~S.~Li$^{9}$,
S.~Allam$^{6}$,
B.~C.~Conn$^{10}$,
C.~Gallart$^{11,12}$,
R.~A.~Gruendl$^{13,14}$,
D.~J.~James$^{15}$,
M.~D.~Johnson$^{14}$,
K.~Kuehn$^{16}$,
N.~Kuropatkin$^{9}$,
N.~F.~Martin$^{17,18}$,
D.~Martinez-Delgado$^{19}$,
D.~L.~Nidever$^{7}$,
N.~E.~D.~No\"el$^{20}$,
J.~D.~Simon$^{21}$,
G.~S.~Stringfellow$^{22}$,
D.~L.~Tucker$^{6}$
}
\vspace{0.4cm}
\\
\parbox{\textwidth}{
$^{1}$ Institute of Astronomy, University of Cambridge, Madingley Road, Cambridge CB3 0HA, UK\\
$^{2}$ Institute of Astronomy and Astrophysics, Academia Sinica, P.O. Box 23-141, Taipei 10617,Taiwan\\
$^{3}$ Center for Computational Astrophysics, Flatiron Institute, 162 5th Avenue, New York, NY 10010, USA\\
$^{4}$ McWilliams Center for Cosmology, Department of Physics, Carnegie Mellon University, 5000 Forbes Avenue, Pittsburgh, PA 15213, USA\\
$^{5}$ Large Synoptic Survey Telescope, 950 North Cherry Avenue, Tucson, AZ 85721, USA\\
$^{6}$ Fermi National Accelerator Laboratory, P.O. Box 500, Batavia, IL 60510, USA\\
$^{7}$ National Optical Astronomy Observatory, 950 N. Cherry Ave., Tucson, AZ 85719, USA\\
$^{8}$ Cerro Tololo Inter-American Observatory, National Optical Astronomy Observatory, Casilla 603, La Serena, Chile\\
$^{9}$ Fermi National Accelerator Laboratory, P.O.\ Box 500, Batavia, IL 60510, USA\\
$^{10}$ Research School of Astronomy \& Astrophysics, Mount Stromlo Observatory, Cotter Road, Weston Creek, ACT 2611, Australia\\
$^{11}$ Instituto de Astrof\'{i}sica de Canarias, La Laguna, Tenerife, Spain\\
$^{12}$ Departamento de Astrof\'{i}sica, Universidad de La Laguna, Tenerife, Spain\\
$^{13}$ Department of Astronomy, University of Illinois, 1002 W. Green Street, Urbana, IL 61801, USA\\
$^{14}$ National Center for Supercomputing Applications, 1205 West Clark St., Urbana, IL 61801, USA\\
$^{15}$ Event Horizon Telescope, Harvard-Smithsonian Center for Astrophysics, MS-42, 60 Garden Street, Cambridge, MA 02138, USA\\
$^{16}$ Australian Astronomical Observatory, North Ryde, NSW 2113, Australia\\
$^{17}$ Observatoire astronomique de Strasbourg, Universit\'e de Strasbourg, CNRS, UMR 7550, 11 rue de l'Universit\'e, F-67000 Strasbourg, France\\
$^{18}$ Max-Planck-Institut f\"{u}r Astronomie, K\"{o}nigstuhl 17, D-69117 Heidelberg, Germany\\
$^{19}$ Astronomisches Rechen-Institut, Zentrum fur Astronomie, Universitat Heidelberg, Monchhofstr. 12-14, 69120 Heidelberg, Germany\\
$^{20}$ Department of Physics, University of Surrey, Guildford GU2 7XH, UK\\
$^{21}$ Observatories of the Carnegie Institution for Science, 813 Santa Barbara St., Pasadena, CA 91101, USA\\
$^{22}$ Center for Astrophysics and Space Astronomy, University of Colorado, 389 UCB, Boulder, CO 80309-0389, USA\\
}
}

\begin{document}

%\date{24th Jan 2017}
\date{\today}

\pagerange{\pageref{firstpage}--\pageref{lastpage}} \pubyear{2017}

\maketitle

\label{firstpage}

\begin{abstract}

We report the discovery of two ultra-faint satellites in the vicinity of the
Large Magellanic Cloud (LMC) in data from the Magellanic Satellites Survey (\maglites). Situated 18$^{\circ}$ ($\sim 20$ kpc) from
the LMC and separated from each other by only $18\arcmin$, Carina~II and III form
an intriguing pair. By simultaneously modeling the spatial and the color-magnitude stellar distributions, we find that both Carina~II and Carina~III are likely dwarf galaxies, although this is less clear for Carina~III.
There are in fact several obvious differences between the two satellites.
While both are well described by an old and metal poor population, Carina~II is
located at $\sim 36$ kpc from the Sun, with $M_V\sim-4.5$ and $r_h\sim 90$ pc, and it is further confirmed by the discovery of 3 RR Lyrae at the right distance.
In contrast, Carina~III is much more elongated, measured to be fainter
($M_V\sim-2.4$), significantly more compact ($r_h\sim30$ pc), and closer to
the Sun, at $\sim 28$ kpc, placing it only 8 kpc away from Car~II. Together with several other systems detected by the
Dark Energy Camera, Carina~II and III form a strongly anisotropic cloud of
satellites in the vicinity of the Magellanic Clouds.

%We present the spectroscopic follow-up data and discuss the satellites' associationwith the LMC and each other in a companion paper.

\end{abstract}

\begin{keywords}
Galaxy: halo, galaxies: dwarf, Magellanic Clouds, Local Group
\end{keywords}

\section{Introduction}\label{sec:INTRO}

%The ``street-light effect'' or the ``drunkard's search'' is an
%allegory for the misguided pursuit of knowledge: a hunt for clues where
%the task is easiest rather than where the truth is most likely. As discussed
%by a medieval sufi, Hodja Nasreddin, generally this approach will lead to an
%observational bias and thus, should be avoided or
%mitigated. Curiously, however, in the long-standing quest for Galactic
%dwarf satellites, the one easy place to look had been ignored until
%recently. The area around the Magellanic Clouds subtends a small
%fraction of the sky and, according to the $\Lambda$CDM cosmology,
%should be littered with small satellites \citep[see,
%  e.g.,][]{Wetzel2015}. Yet, the latest discovery of a large number
%dwarfs next to the Clouds within the Dark Energy Survey (DES) data
%\citep[][]{Koposov2015,Bechtol2015,DrlicaWagner2015,Hor2} appears to
%have come serendipitously.
%{\bf ADW: See notes on Google doc.}

Over the past several decades, the quest for ultra-faint Milky Way satellite
galaxies has been driven by large, multi-purpose digital sky surveys. The
Sloan Digital Sky Survey \citep[SDSS;][]{York2000} and the Dark Energy Survey
\citep[DES;][]{DES2005} have greatly advanced our understanding of the 
ultra-faint galaxy population; however, neither survey was targeted with this
science goal in mind. Thus, one of the most promising regions for finding
ultra-faint galaxies has remained largely unexplored. The region around the
Magellanic Clouds is expected to be littered with small satellites as a byproduct of hierarchical galaxy formation
\citep[e.g.,][]{Springel2008,DOnghia2008,Nichols2011,Sales2011,Wetzel2015}.
Indeed, the recent discovery of a large number of ultra-faint galaxies in DES
data near the Clouds has provided strong support for this hypothesis
\citep[][]{Koposov2015,Bechtol2015,DrlicaWagner2015,Hor2}.

The peculiar alignment of (some of) the DES satellites with the Large
and Small Magellanic Clouds (LMC and SMC) has been scrutinized in
several studies. As first pointed out by \citet{Koposov2015},
under the assumption of isotropy, an overdensity of 9 objects around
the LMC --- as revealed in the DES Year 1 imaging --- is unlikely to
occur by chance. This conclusion is echoed by
\citet{DrlicaWagner2015}, who support their conclusion with the discovery of
8 additional satellites in an adjacent part of the
sky. \citet{Deason2015} study the accretion of LMC analogs in a suite
of cosmological zoom-in simulations of a Milky Way-mass halo and point
out that an excess of satellites around the Clouds would imply 
recent accretion of the Magellanic system onto the Galaxy. A different
approach is taken by \citet[][J16 hereafter]{Jethwa2016}, who model the
distribution of objects within the DES footprint as a mixture of the
virialized Milky Way population and the accreted satellites of
Magellanic origin. Instead of relying on a small number of
cosmological simulations with satellites broadly resembling the LMC
\citep[e.g.,][]{Sales2011,Deason2015,Sales2017}, they follow 
\citet{Nichols2011} to create a framework within which the observed
properties of the Clouds are reproduced. The model includes the
response of the Galaxy to the in-fall of its most massive satellites \citep[e.g.,][]{Gomez2015}, as well as the dynamical friction experienced by
the LMC and SMC \citep[e.g.,][]{Kallivayalil2013}. By
accounting for selection biases, marginalising over the
observational errors and, most importantly, by using a fully
probabilistic approach, J16 demonstrate that between 
one-third and one-half of the dwarfs discovered in the DES could have been delivered by
the LMC.

\begin{figure}
    \includegraphics[width=0.98\columnwidth]{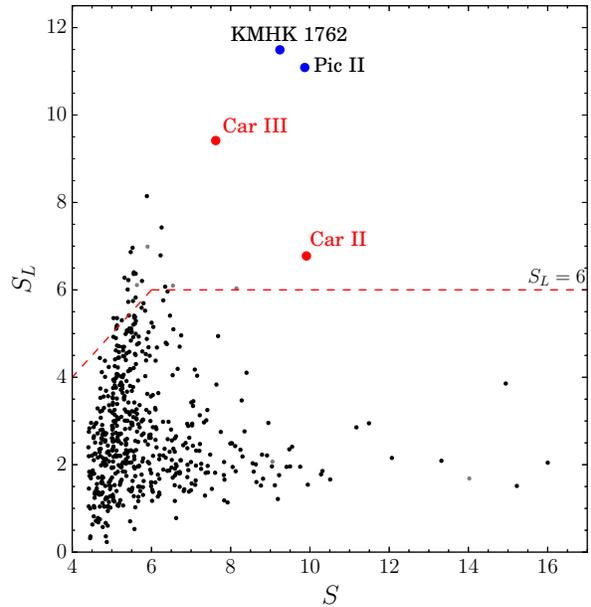}
    \caption{Stellar overdensity significance versus ``local''
      significance (see main text for details) of MagLiteS candidate
      detections. Only stars with $|b|>12\degr$ and distance to the LMC
      greater than $10\degr$ are used. Blue filled circles show the
      locations of two known objects in the field of view: Pictor~II
      and KMHK~1762. In this part of the sky, there are also several
      star clusters: NGC~2808, IC~4499 and E3. However, these are
      detected with significances in excess of 20 and are omitted in
      the plot. Other detected overdensities are shown as black points, the
      new discoveries, Car~II and III are shown as red filled
      circles. There are also a small number of detections associated
      with data artifacts that are shown in light gray. The red
      dashed line marks the $S_L=6$ threshold. Objects with high $S$ but
      low $S_L$ are likely associated with areas in which the variance
      was underestimated. The region above the red line offers a clean
      candidate selection.}\label{fig:svsl}
\end{figure}
\begin{figure*}
  \includegraphics[width=\textwidth]{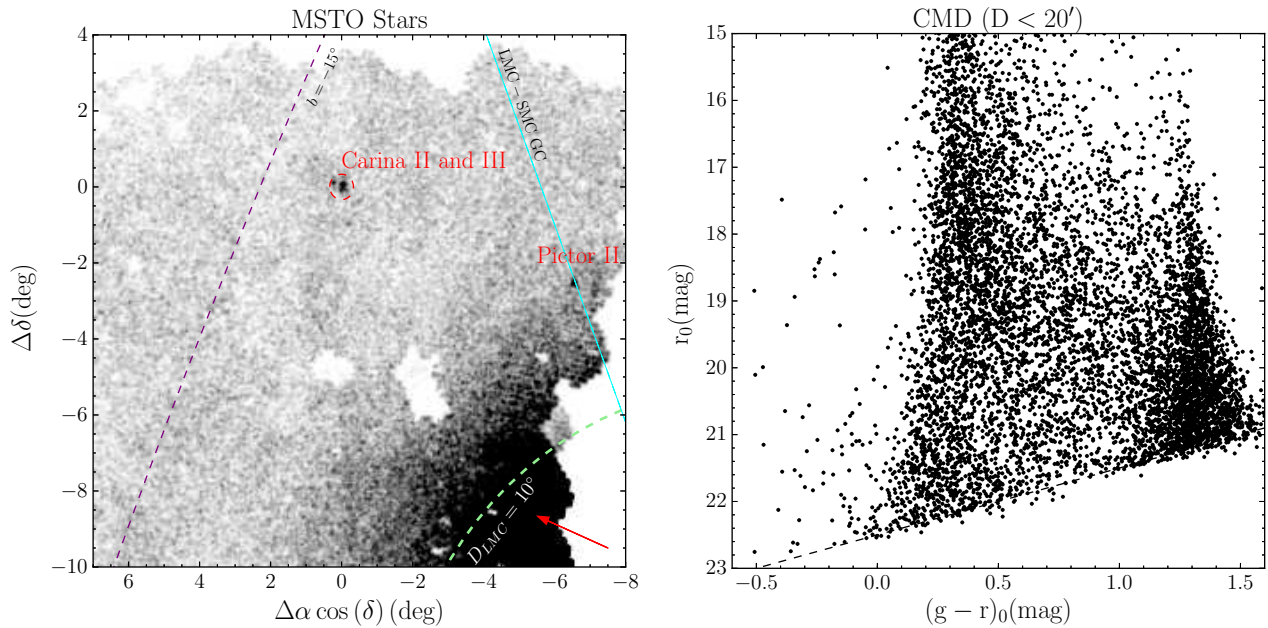}
    \caption{Distribution of stars in the area around Carina~II and III
      in the MagLiteS survey data. {\it Left:} Spatial distribution of
      the MSTO stars at $\sim 35$ kpc, where an overdense region
      corresponding to Car~II and III can be easily seen near the center
      of the figure. The compact overdensity near the right edge of
      the panel is the Pic~II satellite.  The constant Galactic
      latitude line ($b=-15\degr$) is shown in purple, a circle with
      10$\degr$ radius centered on the LMC is shown as a dashed green line. The
      Cloud's proper motion vector shown as a red arrow. The line
      connecting the LMC and SMC is shown in cyan. {\it Right:} CMD of the
      stars within 20\arcmin\ of the center of the Carinas. A sequence
      of BHB stars is clearly visible at $r\sim18.5$ as well as hints
      of a corresponding MSTO at $r\sim21.5$.}\label{fig:rawdata}
\end{figure*}

According to the model of J16, before accretion onto the Galaxy, the LMC
could have harbored as many as $\sim$70 faint dwarf
satellites. Because the Clouds have only recently been accreted onto the Milky
Way, many of their stripped sub-systems have not had much time to
scatter across the sky. As J16 show, even if they have already
been tidally stripped from the LMC, many of the satellites with Magellanic origins
are still to be found in its vicinity. Given that a difference in
energy and angular momentum exists between the LMC and its satellites,
some time after gravitational release the latter will have moved along the
progenitor's orbit, thus populating leading and trailing arms of tidal
debris. The extent and the overall density of the clouds of stripped
satellites around the LMC and the SMC can be used to place stringent
constraints on the masses of the Magellanic Clouds as well as their
orbital history in the Milky Way potential. Additionally, the census of the
Magellanic satellites would enable a unique approach to deciphering
low-mass structure emergence, perhaps shedding unexpected new light on
the nature of the so-called ultra-faint dwarfs \citep[see,
  e.g.,][]{DOnghia2008,Belokurov2013}.

It is with these ideas in mind that the Magellanic SatelLites Survey (MagLiteS) was
conceived. MagLiteS covers
a previously unexplored region of the Magellanic neighborhood
outside of the DES footprint. The survey uses the Dark Energy Camera \citep[DECam;][]{Flaugher:2015} on the Blanco 4m telescope at
Cerro Tololo Inter-American Observatory in Chile to map $\sim 1200$~deg$^{2}$ near the south
celestial pole.  In this context, \citet{Drlica-Wagner2016}
announced the discovery of the Pictor II satellite in Run 1 of MagLiteS, comprising roughly one-quarter of the expected MagLiteS data set. Note that according to the best-fit model
reported by J16, once the survey is completed, a
total of $\sim$5 satellites are expected, of which $\sim4$ should be from the
Magellanic Clouds. 

In this paper we present the discovery of two additional satellites in the
imaging procured by MagLiteS: Carina~II and III. Curiously, these two new
ultra-faint objects form a tight pair on the sky, sitting within $18\arcmin$
of each other. The pair also appears relatively close along the line-of-sight:
their 3D separation is $\sim$8 kpc, while the two are $\sim$30 kpc from the
Milky Way and $\sim$18 kpc from the LMC. Thus, it is reasonable to ask whether
there is or was any physical association between them, and/or with the LMC.
Naturally, before any link between the galaxies is claimed, spectroscopic
follow-up is required. Additionally, we can use the J16 predictions for the
satellite's line-of-sight velocity to address their association to the
Magellanic Clouds.

Note, however, that despite the rigour and the complexity of the
analysis presented in J16, they fail to reproduce the
details of the DES satellite distribution. Broadly speaking, the
spatial density model does not describe the data well on angular
scales of $\sim10^{\circ}$. The mismatch between the data and the
model is largest at lower Magellanic latitudes, or $B_{\rm MS}$, as
defined in the coordinate system linked to the gaseous Magellanic
Stream \citep[see][]{Nidever2008}. On closer inspection, it appears
that a substantial number of DES satellites are arranged in what
resembles a planar configuration. This structure contains seven DES
satellites as well as the LMC and the SMC. The thickness of this
slab-like distribution is $<3$ kpc, but the extent is 90
kpc. J16 estimate that there is a 5$\%$ probability for
this arrangement to happen by chance. Remarkably, all three recent
discoveries by MagLiteS, i.e., Pic~II 
\citep{Drlica-Wagner2016} and Carina~II and III reported here, lie very
close to the plane uncovered by J16.

Our findings on Carina~II and III are presented in two separate papers concentrating on the photometric and spectroscopic analysis, respectively.
This work (Paper~I) is organized as follows: in Section~\ref{sec:Discovery} we describe the MagLiteS data and give the details of the discovery. 
Section~\ref{sec:TimeSeries} describes additional observations acquired with DECam and the time-series analysis of these data.
Section~\ref{sec:Structural} presents the structural and
stellar population modeling of the two new satellites. Finally,
Section~\ref{sec:conclusions} discusses the possible origin of the
Car~II and III pair.
The spectroscopic characterization of Carina~II and III is presented in \citet{carinali}.

\section{Discovery}\label{sec:Discovery}

\subsection{MagLiteS Data}

In this paper, we use $g$ and $r$-band data from the first \maglites observing run (R1) taken over six half-nights between 10 February 2016 and 15 February 2016.
These data were reduced and processed by the DES Data Management system using the pipeline developed for the year-three annual reprocessing of the DES data \citep{Sevilla:2011,Mohr:2012,Morganson:2017}.
Source detection and photometry were performed on a per exposure basis using the \code{PSFex} and \code{SExtractor} routines \citep{Bertin:2011, Bertin:1996}.
Astrometric calibration was performed against the UCAC-4 catalog \citep{Zacharias:2013} using \code{SCAMP} \citep{Bertin:2006}.
The \code{SExtractor} source detection threshold was set to detect sources with  $S/N \gtrsim 5$.
Photometric fluxes and magnitudes refer to the \code{SExtractor} PSF model fit.
The median $10 \sigma$ limiting depth of \maglites is $\gtrsim23$ mag in both bands, which is roughly comparable to the first two years of imaging by DES \citep{DrlicaWagner2015}.

Cross-Matched catalogs were assembled by performing a $1"$ match on objects detected in individual exposures.
Stellar objects were selected based on the \code{spread\_model} quantity: $|\code{wavg\_spread\_model\_r}| < 0.003 + \code{spreaderr\_model\_r}$ \citep[see e.g.,][ for details]{DrlicaWagner2015,Koposov2015}.

Photometric calibration was performed by matching stars to the APASS catalog on a CCD-by-CCD basis \citep{DrlicaWagner2015}.
Extinction from interstellar dust was calculated for each object from a bilinear interpolation of the extinction maps of \citet{SFD}.
We followed the procedure of \citet{Schlafly2011} to calculate reddening, assuming $R_V = 3.1$; however, in contrast to \citet{Schlafly2011}, we used a set of $A_b/E(B-V)$ coefficients derived by DES for the $g$ and $r$ bands: $A_g/E(B-V) = 3.683$ and $A_r/E(B-V) = 2.605$.

\subsection{Discovery}

\begin{table*}
    \tiny
        \centering
        \caption{Coordinates, period, amplitudes, mean magnitudes in $gri$, and distance for the 5 RR Lyrae stars in the DECam field.}
        \label{tab-rrls}
        \begin{tabular}{lccccccccccrrc} 
        \hline
ID & RA     & DEC    & Type & Period & $\Delta_g$ & $g$    & $\Delta_r$ & $r$     & $\Delta_i$ & $i$      & $d_{\rm Car2}$ & $d_H$ & Var \\ 
   & (deg) & (deg)   &      & (d)    & (mag)      & (mag)  & (mag)      & (mag)   & (mag)      & (mag)    & ($\arcmin$)       & (kpc) & \\
                \hline
MagLiteS\_J073637.00-580114.4 & 114.15416 & -58.02068  & ab & 0.6424 & 0.96  & 18.88  & 0.67 & 18.57 & 0.56  & 18.42 &  2.1  & $38.1 \pm 0.4  $ & CarII-V1 \\ 
MagLiteS\_J073704.16-575334.4 & 114.26732 & -57.89288  & c  & 0.3654 & 0.58  & 19.56  & 0.39 & 19.31 & 0.30  & 19.21 &  7.3  & $47.6 \pm 0.6$ & \\
MagLiteS\_J073645.85-575154.1 & 114.19105 & -57.86502  & c  & 0.4079 & 0.50  & 18.81  & 0.35 & 18.55 & 0.28  & 18.44 &  7.7  & $37.3 \pm 0.4$ & CarII-V2 \\
MagLiteS\_J073509.12-575714.8 & 113.78799 & -57.95410  & ab & 0.7051 & 0.89  & 18.70  & 0.64 & 18.39 & 0.52  & 18.27 &  10.9 & $36.7 \pm 0.4$ & CarII-V3 \\
MagLiteS\_J073843.91-582645.9 & 114.68295 & -58.44607  & ab & 0.6243 & 0.79  & 16.62  & 0.60 & 16.28 & 0.49  & 16.12 &  32.6 & $12.0 \pm 0.1$ & \\
                \hline
        \end{tabular}
\end{table*}

Carina~II and III were discovered by applying a version of the satellite
search algorithm described in \citet{2016MNRAS.463..712T} to the
MagLiteS data. In short, the algorithm computes the local stellar density and
an estimate of the local background by convolving the star count map
with two two-dimensional Gaussian kernels. A large kernel with
$\sigma_o=60$\arcmin\ is used for the background estimation, which is then subtracted from a
suite of 5 different small kernels with $\sigma_i=
1\arcmin,2\arcmin,5\arcmin,8\arcmin$ or 10\arcmin\ used for the
local density estimation. Different $\sigma_i$ are used to detect
satellites of different apparent sizes, since the search is more
sensitive to overdensities with extents similar to that of the kernel
\citep{Koposov2008}. Prior to spatial convolution, we select stars using
an isochrone mask that picks out a specific stellar population at a
particular distance. In the analysis presented here, we use a single
PARSEC isochrone \citep{Bressan2012} with $[{\rm Fe/H}]=-2$ and an age of
$\sim 12$\,Gyr. To detect stellar systems at different distances, the
isochrone is shifted to 36 different distance moduli between $16 \leq m-M \leq 23$, corresponding to satellites with heliocentric distances between  $\sim16 \leq D \leq 400$\,kpc. The significance,
$S$, of an overdensity is then estimated by comparing the results of
the convolution with the expected variance.

In the current implementation, we also derive a local significance to calibrate candidate identification in areas where the variance
is underestimated. This is particularly relevant for the MagLiteS
dataset, given its proximity to the LMC - a portion of the sky with a
highly variable background. Rapid fluctuations in the stellar density
field break the assumptions underlying the overdensity search as
described above, and bias the estimates of the local background and
its variance. To avoid a large number of false positives in areas
close to the LMC, we introduce the local significance, $S_L$, based on the properties
of $S$ around an overdensity:

\begin{equation}
  S_L=\frac{S(0)-\left<S_{d<\sigma_o}\right>}{\sqrt{\rm{Var}\left(S_{d<\sigma_o}\right)}},
\end{equation}

\noindent where $S(0)$ is the significance at the center of an
overdensity, and $S_{d<\sigma_o}$ are the significances for all pixels
within $\sigma_o$ from the center. In areas where the variance is
underestimated, $S$ is overestimated and then
$\left<S_{d<\sigma_o}\right>\gg 0$, which means $S_L\ll S$. This
allows us to cull false positives by simply selecting
overdensities that have both large $S$ and large $S_L$.

Figure~\ref{fig:svsl} displays the values of $S$ and $S_L$ for all
overdensities that pass our minimum significance cut and are located
sufficiently far from both the Galactic plane ($|b|>12\degr$) and the
LMC ($D_{LMC}>10\degr$). Previously known objects are shown in blue,
while unidentified candidates are shown as black points. In gray we show a
small number of overdensities associated with obvious data artifacts
that had to be removed manually\footnotemark. The two newly discovered
objects that are the focus of this paper are shown in red. These
systems are both located in the constellation of Carina, which already
hosts a classical dwarf spheroidal galaxy \citep[][]{carinadisc}, and therefore are assigned the names Carina~II and Carina~III. In the list
from which we pick trustworthy satellite candidates with $S_L>6$,
Car~II appears as the most significant detection with $S=9.9$, and
Car~III as the second most significant detection, with $S=7.6$. On the
sky, the two are very close to each other, only $\sim18\arcmin$ apart,
but are detected as two distinct overdensities by our algorithm. Other candidates above the $S_L=6$ line are not obviously spurious nor obviously real, so more analysis/data is needed to confirm or discard them.

%Car~II and Car~III were also identified as statistically significant
%overdensities by the maximum-likelihood search algorithm applied in \citet
%{Drlica-Wagner2016}. However, the small separation between the two led to a
%complex spatial morphology, and the pair were set aside for future
%investigation.

The left panel of Figure~\ref{fig:rawdata} shows the density distribution
of the main sequence turn-off (MSTO) stars in the area around
Carinas. The MSTO stars are selected with an isochrone mask based on
an old and metal poor population at the distance of 35\,kpc and $r>21$. The map highlights the almost continuous
MagLiteS coverage of the area and the stability of its photometry. Note that the $r>21$ cut is just to select MSTO stars, and do not represent the limiting magnitude of \maglites, which is closer to $\sim23$ magnitudes. The
outskirts of the LMC's disc are clearly visible, extending beyond 10
degrees from the Cloud's center. In fact, a diffuse stellar cloud
associated with the LMC can be seen stretching as far as
$\sim15^{\circ}$ and engulfing the recently discovered Pictor II \citep[Pic~II;][]{Drlica-Wagner2016}. Carina~II and III stand out
dramatically in the density map at an angular separation of $\sim18^{\circ}$
from the LMC. As this Figure demonstrates, the Carinas are
$\sim 17^{\circ}$ from the Galactic plane. Despite the close proximity 
to the Galactic disk, the extinction is low and therefore does
not seem to affect the stellar distribution significantly.

The right panel of Figure~\ref{fig:rawdata} presents the colour-magnitude
diagram (CMD) of stars centered on Carina~II. The portion of the CMD with
$r<20$ is dominated by the Galactic disc dwarfs (unsurprisingly, given the low
Galactic latitude of the object), thus concealing the object's Red
Giant Branch (RGB). Nonetheless, the pile-up of stars around the
satellite's turn-off is visible at $(g-r)_0 \sim 0.4$ and $r_0\sim 21.5$. Even
more obvious are the Blue Horizontal Branch (BHB) stars at $(g-r)_0<0$ and
$r\sim 18.5$. Additionally, right under the BHB, a sprinkle of stars that looks like Blue
Straggler can be seen at $(g-r)_0<0$ and $r\sim 21$.

\begin{figure*}
  \begin{center}
    \includegraphics[width=0.8\textwidth]{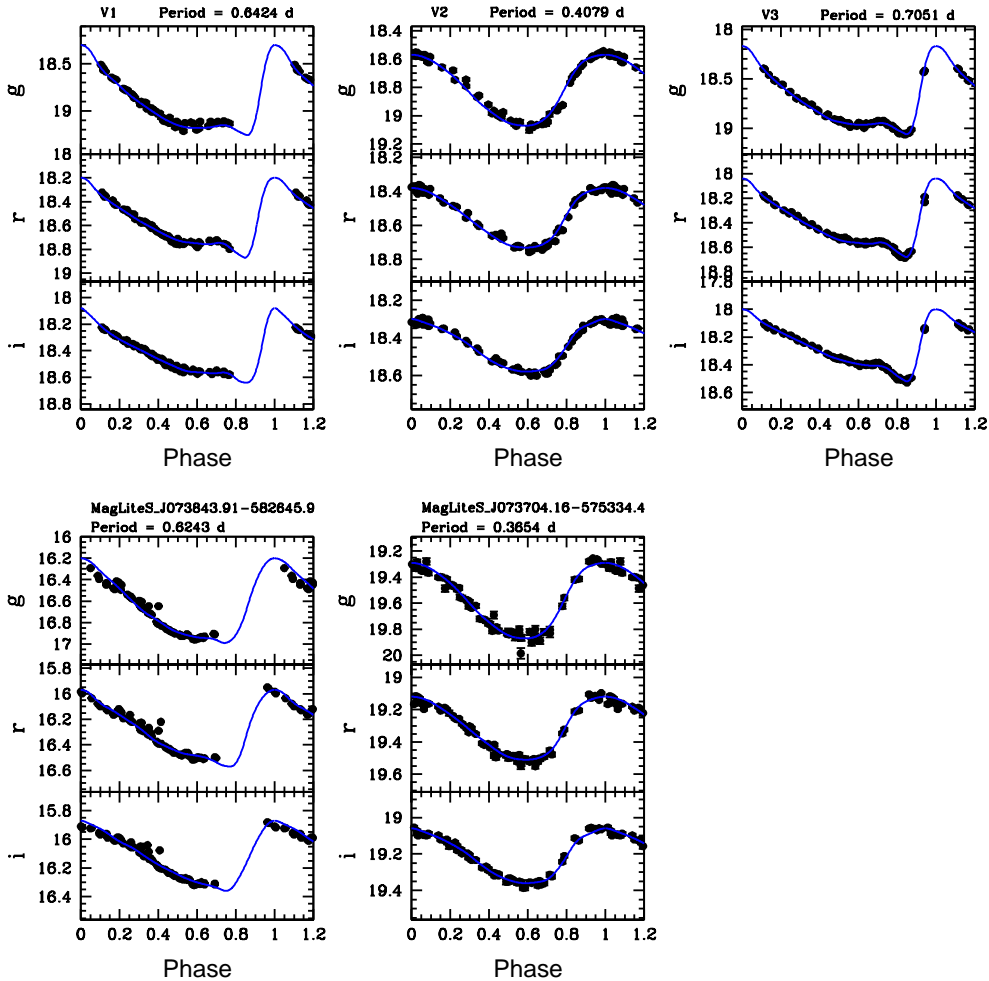}  
    \caption{Light curves of the 5 RR Lyrae stars - Two RRc stars (middle panels) and three RRab stars - found within 0.6 degrees from the center of Carina~II. 
    The three stars in the top row (CarII-V1 to CarII-V3) have mean magnitudes and location consistent
    with being members of Carina~II. The two stars in the bottom row are halo stars along the line of sight, although MagLiteS\_J073704.16-575334.4, located at $\sim 48$ kpc might be associated with the LMC. The solid blue lines are templates fitted to the data from 
    the library of \citet{Sesar2010}.}
    \label{fig-lightcurve}
  \end{center}
\end{figure*}

\footnotetext{These are easy to identify and removed due to their
  rectangular CCD chip-like shapes.}

\section{Follow-up Imaging}\label{sec:TimeSeries}

In addition to the original MagLiteS data, two series of follow-up DECam imaging were obtained of the Carina~II+III field during Blanco 4m Director's Discretionary and engineering time on the nights of 2017 January 17 and February 12, both of them with bright moon. During each of those two nights the Carinas were observed multiple times in $gri$ over a period of $\sim 7$ hours. Several $u$-band exposures were also taken during the periods of time on those nights when the Moon was below the horizon. In addition, several $gri$ exposures were obtained on each of 2017 February 9, 10 and 11. 
In total, we obtained 42 epochs in $u$ and 70 epochs in each $gri$.
Exposure times were 120s in $u$ and 60s in $gri$. Both galaxies were covered by a single DECam field. 
The observing sequence consisted of a $[u]gri$ of a field in which Carina~II was centered on one of the central CCDs of the camera (N4), 
followed immediately by another sequence with offsets in RA and DEC of $60\arcsec$ to cover the gaps between CCDs.

These individual exposures were processed with software equivalent to the standard DES processing pipeline. The astrometric residuals are $\sim 40$ mas rms and the photometric calibration has an accuracy of 2\% rms in $gri$ and 5\% in $u$.  Objects were detected using standard SExtractor (a simple cut on $|\rm spread\_model| < 0.003$ was used to separate stars from extended objects and objects with SExtractor flags $< 4$ were kept allowing close blends in this relatively crowded field at $(l,b) = (270^\circ, -17^\circ)$. Individual exposure and filter catalogs were matched using the {\it STILTS} software package\footnote{http://www.star.bris.ac.uk/$\sim$mbt/stilts} \citep{2006ASPC..351..666T}.  This yielded ugri lightcurves with up to 70 epochs for over 30,000 objects within 0.6 degrees of the center of Carina~II+III.  RR Lyrae stars associated with Carina~II were identified as described below and used to refine the distance to the Carina~II system, along with a number of field eclipsing binaries.   

In addition to the light curve data, deep coadds were constructed from all of the available imaging data. Catalogs from these coadds were extracted and resulted in detection limits approximately half a magnitude deeper in g and r bands ($r \sim 23.5$) than the discovery MagLiteS exposures themselves.

% CMDs on and off the Carinas are shown in Figure~\ref{fig:hess} and were used to refine the best-fit isochrones for Carina~II and III.  The distances derived from isochrone fits to these deeper coadds are in excellent agreement with the estimate based on RR Lyrae members of Carina~II (Tables \ref{tab:Properties} and \ref{tab-rrls}).

\subsection{RR Lyrae Stars}

To search for variable stars we used the time series data in $gri$ only because the $u$ data does not
have an extended time baseline.
We flagged stars as variable if the standard deviation of their magnitudes in each band is $3\sigma$ 
above the distribution for the bulk of the population (which is
non-variable) of similar magnitude. In order to avoid spurious variables, we required that the stars 
were flagged as variable in all 3 bands. The amplitude of variation of RR Lyrae stars is large enough 
in all optical bands that there is no risk of missing one due to this requirement. This selection results in 167 variable stars (0.7\% of the total) for further study.

The selected variable stars were then searched for periodicity using an implementation of the well known \citet{Lafler1965} algorithm that uses the information in all three bands simultaneously \citep[see][]{Vivas2016}. We searched for RR Lyrae stars in the range 0.2 to 0.9 days, and for SX Phe/$\delta$ Scuti  stars in the range 0.01 to 0.2 days. In total we found 51 periodic variable stars in the field, 5 of which we classified as RR Lyrae stars. The remaining 46 stars were different types of eclipsing systems (see Appendix~\ref{sec-othervariables}).No short period variables exemplifying the SX Phe or $\delta$ Scuti stars were found. On the other hand, our cadence is not sensitive to periods larger than 1 day and hence, the Anomalous Cepheid stars range was not fully explored. Nevertheless, we looked at the time series of all variable stars candidates in the region of the CMD where Anomalous Cepheid are located, searching for smooth variations in timescales $>1.0$d, and found none. Light curves in $gri$ for the 5 RR Lyrae stars (3 of type ab and 2 of type c) found in the field are shown in Figure ~\ref{fig-lightcurve}. 

\begin{figure*}
    \includegraphics[width=\textwidth]{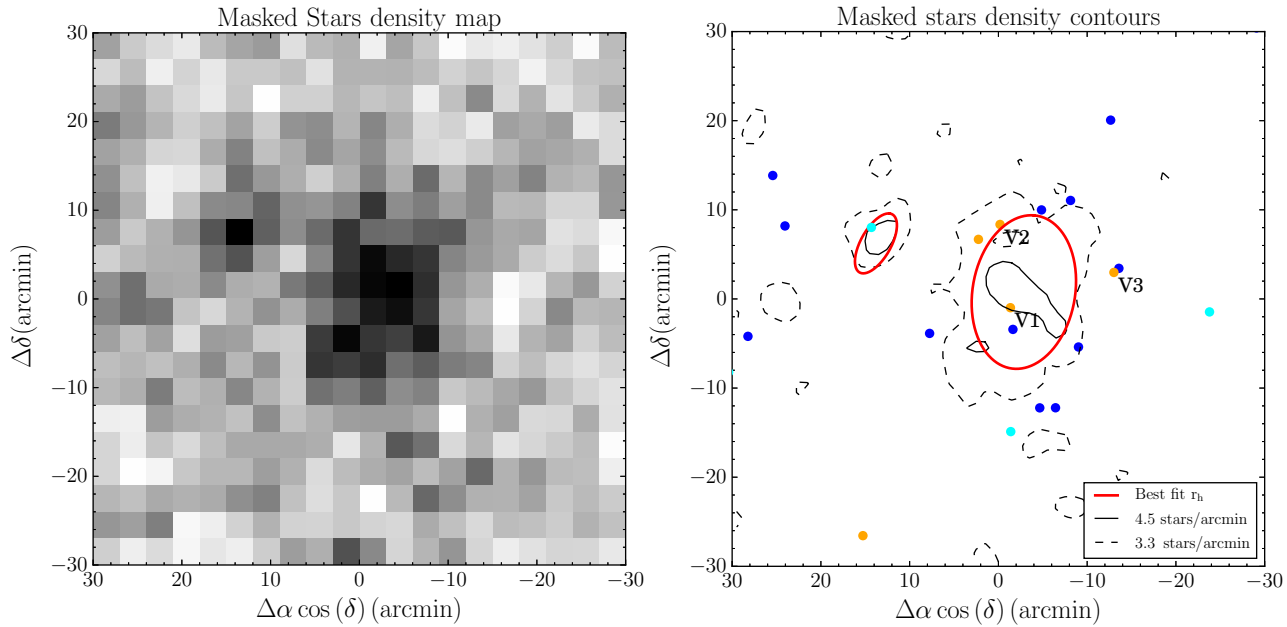}
    \caption{Deep follow-up imaging of the Carina~satellites {\it
        Left:} Density map of stars within a CMD mask defined by the
      best-fit isochrone of Carina~II. {\it Right:} Density contours of
      the same region of the sky with the half-light radius of the
      best-fit models overplotted in red. Stars within
      0.3 magnitudes from the BHB ridge line of Car~II(III) are shown in
      blue(cyan), and RR Lyrae stars found in the field are shown in orange. Both overdensities are clearly seen in the density
      maps and are well reproduced by the
      models.} \label{fig:spatialfit}
\end{figure*}

We obtained mean magnitudes of the RR Lyrae stars by fitting templates from the library by \citet{Sesar2010} which is based on RR Lyrae stars in SDSS Stripe 82. 
The light curves were integrated in intensity units and the mean value transformed back to magnitudes. By using a template we overcome any possible bias in the mean magnitude due to uneven sampling of the light curve, as is the case for several of our stars (Figure~\ref{fig-lightcurve}).  In Table~\ref{tab-rrls} we provide coordinates, periods, amplitudes and mean magnitudes in each band for the RR Lyrae stars. The last three columns in Table~\ref{tab-rrls} contain the separation in the sky (in arcmin) from the center of Carina~II, the heliocentric distance which was calculated as described in \S~\ref{sec-distance}, and a variable designation (CarII-V1 to CarII-V3) for the 
stars which are members of Carina~II.

From the spatial distribution of the RR Lyrae stars in the region (see Figure~\ref{fig:spatialfit}) it is clear that all but one of the RR Lyrae stars are close enough, $< 11\arcmin$, to the center of Carina~II to suspect a relationship. However, the location of these stars in the CMD (Figure~\ref{fig:hess}) confirms that only 3 of them (MagLiteS\_J073637.00-580114.4, MagLiteS\_J073645.85-575154.1 and MagLiteS\_J073509.12-575714.8) are at the level of the horizontal branch of Carina~II. These stars were designated as CarII-V1 to CarII-V3. Stars CarII-V1 and CarII-V2 were
recently confirmed as members of Carina~II based on their radial velocities \citep{carinali}. 
No spectroscopic observations for CarII-V3 are yet available.

Star MagLiteS\_J073704.16-575334.4 is at only $7.3\arcmin$ from the center of Carina~II but it is $\sim 0.8$ mag fainter than the other three stars. Thus, this star, as well as MagLiteS\_J073843.91-582645.9, which is $\sim 33\arcmin$ from the center of Carina~II and $\sim 2$ mags brighter than its HB, are not associated with either of the Carinas. 

None of the RR Lyrae stars are located near Carina~III. If Carina~III is confirmed as a UFD (rather than a stellar cluster), this would be the first of all satellite galaxies --- that have been searched for variable stars --- in which no RR Lyrae stars have been found \citep[see compilations in][]{Baker2015,Vivas2016}.

\subsection{Distance to Carina~II from RR Lyrae Stars \label{sec-distance}}

To calculate the heliocentric distances to the RR Lyrae stars we used the relationships for the absolute magnitude in the SDSS-$i$-band ($i_S$, where the subscript $S$ stands for SDSS), $M_i$, provided by \citet{Caceres2008}, which are based on
theoretical models of these stars in the SDSS bands, and depend on both the period and the metallicity for the star. Since our magnitudes are tied to the DES (AB) photometric system, we first transformed
our $i$ mean magnitudes to $i_S$ using

\begin{eqnarray}
i &=& i_S + 0.014 - 0.214\,(i-z)_S - 0.096\,(i-z)^2_S \\
z &=& z_S + 0.022 - 0.068\,(i-z)_S \nonumber
\end{eqnarray}

These transformation equations contain a colour term with $(i-z)$, which we do not have in our time series observations. Because RR Lyrae
stars have very small dispersion in the mean $(i-z)$ colour distribution \citep{Vivas2017}, we used the mean colour of the RR Lyrae stars in the globular cluster M5 ($(i-z)_0=0.013$ and $-0.006$, for ab and c-type respectively) as provided in \citet{Vivas2017}. For the stars suspected to be members
of Carina~II, we assumed a metallicity of [Fe/H]$=-2.4$ based on the spectroscopy recently obtained by \citet{carinali} for this galaxy. We also
assumed [$\alpha$/Fe]$=0.2$, which is typical of ultra-faint galaxies \citep{Pritzl2005}. For the two suspected halo stars, we assumed 
[Fe/H]$=-1.65$ and [$\alpha$/Fe]$=0.3$. Individual heliocentric distances are provided in Table~\ref{tab-rrls}.

Thus, the distance to Carina~II based on its 3 RR Lyrae stars is $37.4 \pm 0.4$ kpc, in agreement with the structural parameters derived
from the CMD fitting described in the next section. Star MagLiteS\_J073843.91-582645.9 is a foreground halo star at $12.0 \pm 0.1$ kpc from the Sun, while MagLiteS\_J073704.16-575334.4 is behind Carina~II,
at $47.6 \pm 0.6$ kpc. The latter is located exactly at the distance at which material from the LMC would be expected \citep{Munoz2006}. Spectroscopy will be needed to confirm if this distant RR Lyrae star is indeed associated with the LMC.

\begin{table}
    \caption{Properties of Car~II and Car~III}
    \centering
    \label{tab:Properties}
    \begin{tabular}{@{}lrrl}
        \hline
        Property               & Carina~II              & Carina~III                         & Unit\\
        \hline
        $\alpha({\rm J2000})$  & $114.1066 \pm 0.0070$  & $114.6298 \pm 0.0060$ & deg \\
        $\delta({\rm J2000})$  & $-57.9991 \pm 0.0100$  & $-57.8997 \pm 0.0080$ & deg \\
        $(m-M)$                & $17.79 \pm 0.05$       & $17.22 \pm 0.10$     & mag\\
        $D_\odot$              & $36.2 \pm 0.6$         & $27.8 \pm 0.6$       & kpc\\
        $r_{h}$                & $8.69 \pm 0.75$        & $3.75 \pm 1.00$         & arcmin\\
        $r_{h}$                & $91 \pm 8$            & $30 \pm 9$           & pc\\
        1$-$b/a                & $0.34 \pm 0.07$       & $0.55 \pm 0.18$      & \\
        PA                     & $170 \pm 9$           & $150 \pm 14$         & deg\\
        $M_V$                  & $-4.5 \pm 0.1$        & $-2.4 \pm 0.2$       & mag\\
        $[$Fe/H$]$   & $-1.8\pm0.1$          & $-1.8\pm0.2$         & dex\\
        Age          & $9.9\pm0.4$           & $9.7\pm0.8$          & Gyr\\
        \hline
    \end{tabular}
\end{table}

\subsection{Properties of the Carina~II RR Lyrae stars}

The pulsational properties of RR Lyrae stars in satellite galaxies are useful to understand
the role of these systems in the formation of large galaxies like the Milky Way \citep{Clementini2014,zinn14,Fiorentino2015,Vivas2016}.
In particular, the mean period of the fundamental mode RR Lyrae stars (ab-type) can be used to classify the stellar system in 
Oosterhoof (Oo) groups \citep{Oosterhoff1939,Catelan2015}. 
With only a couple of exceptions \citep[Canes Venatici I and Ursa Major I,][]{Clementini2014} that belong to an intermediate Oo group, 
all other ultra-faint dwarf (UFD) galaxies have been classified as Oo II \citep[see compilation in][]{Vivas2016}. Carina~II follows that trend.
Although it has only 2 ab-type RR Lyrae, their mean period is 0.67d, close to the nominal 0.65d of the OoII group \citep{Catelan2015}.

With 3 RR Lyrae stars, Carina~II resembles Leo V, a UFD with similar absolute magnitude, $M_V=-4.4$, and
containing also 3 RR Lyrae stars \citep{Medinasubmitted}. Ursa Major II and Canes Venatici II which 
also have similar brightness,
$M_V=-4.0$ and $-4.6$ respectively \citep{Sand2012}, have 1 and 2 RR Lyrae stars
\citep{Dallora2012,Greco2008}. Thus, the production of RR Lyrae stars in Carina~II follows the trend of galaxies of 
similar brightness in the Milky Way system.

\section{Structural Parameters}\label{sec:Structural}

We measure the properties of Carina~II and III by modeling the
magnitudes, colours, and positions of stars extracted from the deep coadded images (see above) in an area $\sim 1^{\circ} \times 1^{\circ}$
centered on Car~II \citep[see e.g.][for a similar
  approach]{Martin2008,Koposov2015,2016MNRAS.463..712T}. To avoid
density variations caused by spatially varying incompleteness as well
as to avoid the contamination due to misclassified galaxies, we only
consider stars brighter than $23$ in $g$ and $r$. We also correct for extinction using the \citet{SFD} maps, but used updated coefficients for the DES filters, namely, $A_g/E(B-V) = 3.184$ and $A_r/E(B-V) = 2.13$. Additionally, given
the close proximity of the two satellites to each other, we model
their stellar components simultaneously. Therefore, our model
probability density function (PDF) describing the positions, colours, and magnitudes of the stellar data has three components,
namely, the background plus the two satellites:

\begin{equation}
  P(\Phi|\Theta)=\prod\limits_i \left(
f_1P_s(\Phi_i|\Theta_1)+f_2P_s(\Phi_i|\Theta_2)+f_bP_b(\Phi_i|\Theta_b)\right),
\end{equation}

\noindent where the product is over all stars within the modeled
area. $\Phi=(x,y,g-r,r)$ is a shorthand for the observed properties of
the stars, namely, the on sky positions $(x,y)$, their colours $(g-r)$,
and magnitudes $r$. $f_1$ is the fraction of stars in Car~II, $f_2$ the
fraction of stars in Car~III, and $f_b=1-f_1-f_2$ the fraction of stars
in the background. $\Theta$ stands for all of the model parameters,
which are divided between the three components using the same
sub-indexes as the fractions $f$.

\begin{figure*}
    \includegraphics[width=\textwidth]{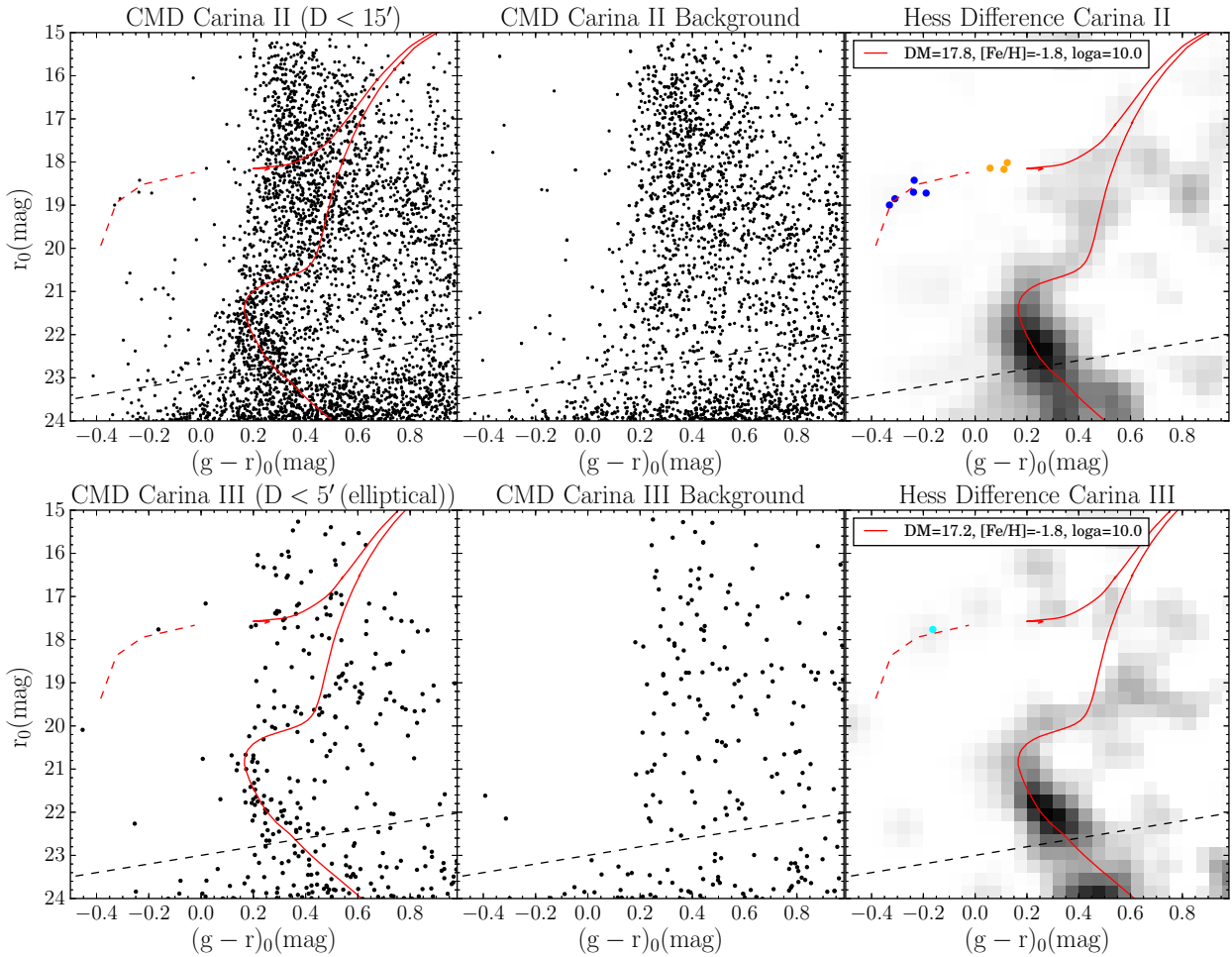}
    \caption{CMD and Hess difference diagram for the Carina~II and III
      systems. Top (bottom) row presents CMD information for Carina~II
      (Carina~III). Left panel gives the CMD within twice the half light
      radius of the satellite, middle panel shows the CMD of the
      foreground stellar population estimated for the same solid angle as used for the
      left panel. Finally, right panel demonstrates the foreground
      subtracted Hess difference diagram. Solid red line shows the
      best-fit isochrone (see Section~\ref{sec:Structural} for
      details), while the red dashed line indicates the M92 BHB
      ridge-line shifted to the distance of each satellite. In blue
      (cyan) we identify individual BHB candidates for Carina~II
      (III), and in orange we show RR Lyrae stars identified at the same distance as Carina~II. Note that the neither the BHB stars nor the RR Lyrae stars were used in the determination
      of the best isochrone and therefore provide an independent
      confirmation of the distance. For both Carina~II and III the Hess
      difference diagrams and the CMDs clearly show prominent MSTO
      features that are absent in the foreground thus confirming that
      Carinas are geniune stellar systems in the Milky Way
      halo. The black dashed line is the $g,r=23$ magnitude limit used in the modeling.} \label{fig:hess}
\end{figure*}

For each satellite, the PDF, $P_s(\Phi|\Theta_{k})$ is composed of the
spatial model, $P_{sp}(x,y|\Theta_{k,sp})$, and the colour-magnitude model,
$P_{cmd}(g-r,r|\Theta_{k,cmd})$. The spatial model for the Carinas is defined as a 2-D
elliptical Plummer sphere:

\begin{equation}
  P_{sp}(x,y|\Theta_{k,s})=\frac{1}{\pi a^2 \left(1-e\right)}\left(1+\frac{\tilde{r}^2}{a^2}\right)^ {-2},
\end{equation}

\noindent where $\tilde{r}^2=\tilde{x}^2+\tilde{y}^2$ is the elliptical radius and

\[
\begin{bmatrix}
  \tilde{x} \\
  \tilde{y}
\end{bmatrix}
=
\begin{bmatrix}
  \cos{\theta}/(1-e) & \sin{\theta}/(1-e)  \\
  -\sin{\theta}      & \cos{\theta}
\end{bmatrix}
\begin{bmatrix}
  x-x_0 \\
  y-y_0
\end{bmatrix},
\]

\noindent where the 5 parameters of the spatial model are: the center
of the Plummer profile $(x_0,y_0)$, the elliptical half-light radius $a$, the
ellipticity $e$, and the positional angle of the major axis $\theta$.

The PDF in colour-magnitude space, $P_{cmd}$, is defined using PARSEC
isochrones \citep{Bressan2012} populated according to the
\citet{Chabrier2003} stellar mass function. Specifically, this PDF
is constructed on a pixel grid in the ($g-r$,$r$) space and the
probabilities of finding a star in each bin given an isochrone are
found by convolving the expected number of stars along the isochrone
track with the corresponding photometric errors. This probability
distribution depends on three parameters: the isochrone age,
metallicity, and the distance modulus. One should be aware, however, that this approach comes with several limitations. In particular, the uncertainties in the modeling of the isochrones, which can give rise to systematic errors \citep[see e.g. ][]{Drlica-Wagner2016} are not considered. Also, the PARSEC isochrones only have metallicities down to -2.1 dex \citep{Bressan2012}, and hence the presented approach cannot account for metallicities lower than this limit. Furthermore, the differences between isochrones become smaller at decreasing metallicities, making it more difficult to distinguish different metallicities at the low-metallicity end. This issue is of particular importance for faint satellites - like the ones presented in this paper - since the main effect of the metallicity to the shape of the isochrone, at fixed age an distance, is to shift the RGB, which is poorly populated in these objects\footnotemark.

\footnotetext{Note that, at fixed age and distance, the difference between the tip of the RGB gets as small as 0.02 in $g-r$ and less than 0.1 magnitudes in $r$ between the two lower metallicity isochrones.}

The PDF for the stellar foreground/background population is also
defined as the product of the spatial component and a colour-magnitude component:
$P_b(\Phi|\Theta_{bg})=P_{bg,sp}(x,y|N_{bg},b_1,b_2)P_{bg,cmd}(g-r,r)$. The
spatial model for the background is defined as a bilinear distribution
of the form:

\begin{equation}
  P_{bg,sp}(x,y|N_{bg},b_1,b_2) = \frac{1}{N_{bg}}(b_1 x + b_2 y + 1),
\end{equation}

\noindent where $b_1$ and $b_2$ define the two-dimensional background
gradient, and $N_{bg}$ is defined so $P_{bg,sp}$ is normalized to one over
the modeled area. The colour-magnitude component of the background does not have
any extra parameters, since it is defined empirically by constructing
a histogram in colour-magnitude space using the stars outside
20\arcmin\ from the center of Car~II, and normalized to unity.

% \begin{figure}
%     \includegraphics[width=0.98\columnwidth]{img/radialprofile.eps}
%     \caption{Azimuthally averaged stellar density profiles for Carina~II (top panel), and  Carina~III (bottom panel). The solid red line shows the combined best fit model while the red dashed lines show individual models for Carina~II and III. The black dashed line shows the half-light radius.\skcom{You cannot really see individual models except the background...}.}\label{fig:radialprofile}
% \end{figure}
% %

%
\begin{figure*}
    \includegraphics[width=\textwidth]{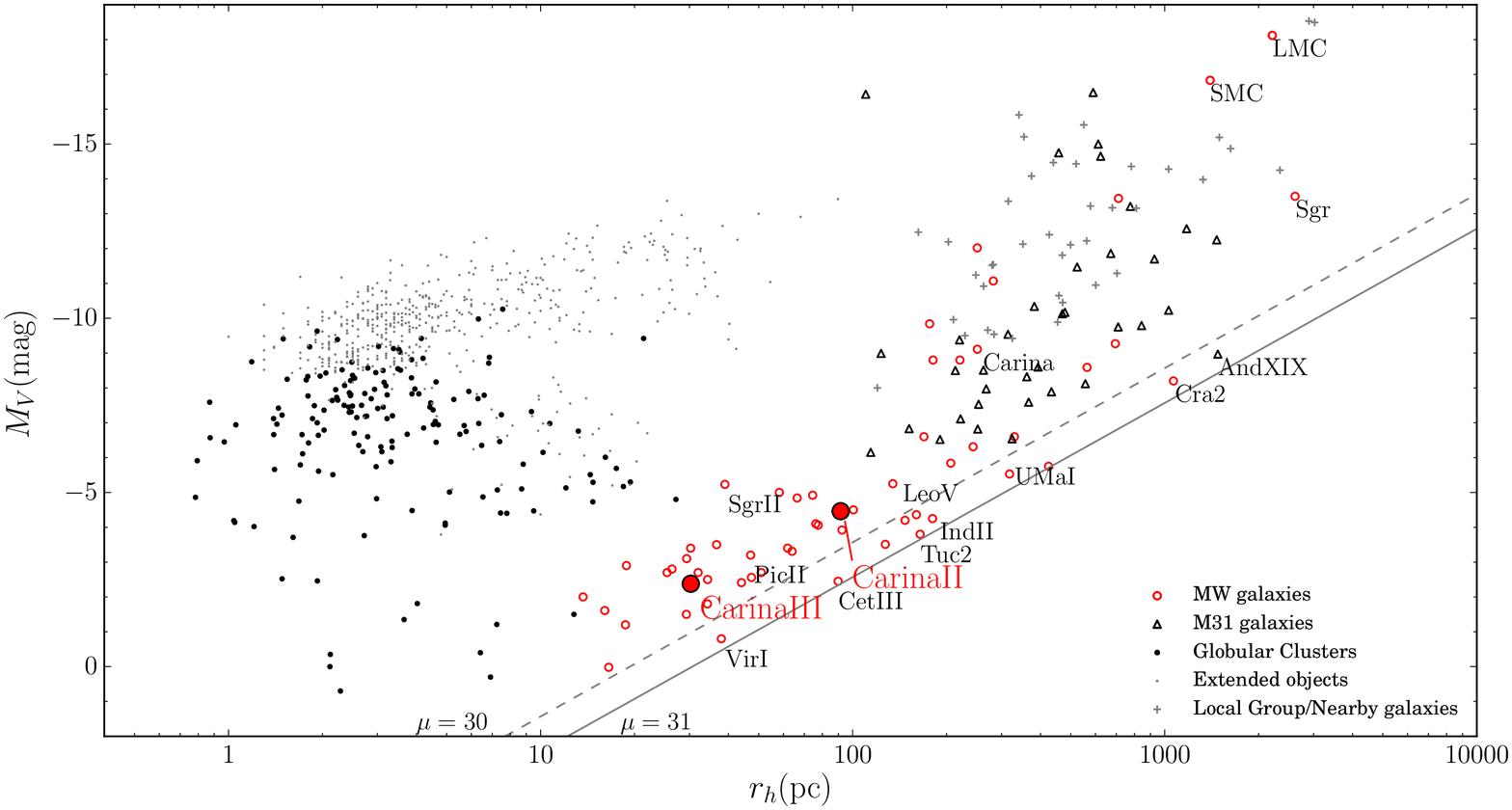}
    \caption{Absolute magnitude versus half-light radius for stellar
      systems in the Local Group. Local galaxies from
      \citet{McConnachie2012} (updated September 2015): dwarf galaxy
      satellites of the Milky Way are shown with red open circles, the
      M31 dwarfs with black unfilled triangles, and other nearby
      galaxies with gray crosses. The positions of Crater~2
      \citep{Torrealba2016}, Aquarius~2 \citep{2016MNRAS.463..712T},
      DESJ0225+0304 \citep{2016arXiv160804033L}, Pic~II
      \citep{2016ApJ...833L...5D}, Virgo~I \citep{2016ApJ...832...21H}
      and Cetus~III \citep{2017arXiv170405977H} are also
      displayed. Black dots are the Milky Way globular clusters
      measurements
      \citep[from ][]{Harris2010,Belokurov2010,Munoz2012,Balbinot2013,Kim2015a,Kim2015b,Kim2016,Laevens2015,Weisz2016,Luque2015,2016arXiv160804033L,2016ApJ...830L..10M,2017arXiv170201122K}
      and gray dots are the extended objects smaller than 100\,pc from
      \citet{Brodie2011}. The black solid (dashed) line corresponds to
      the constant (effective) surface brightness (i.e that within
      half-light radius) of $\mu=31$ (30)
      $\rm{mag}\,\rm{arcsec}^{-2}$. 
      The surface brightness limit of the searches for
      resolved stellar systems in the SDSS \citep{Koposov2008} and
      similar surveys is $\lesssim 30\,\rm{mag\,arcsec}^{-2}$.} \label{fig:mvrh}
\end{figure*}

The full model contains 20 free parameters, 2 for the background and 9 for
each of the satellites. We sample the likelihood with the affine invariant
ensemble sampler \citep{GoodmanWeare2010} implemented in the {\it emcee}
python module \citep{ForemanMackey2013}. We use flat priors over all the
parameters except the spatial size parameter, $a$, of both satellites, for
which we use the uninformative Jeffreys prior $P(a)\propto\frac{1}{a}$. We
measure the best-fit parameters and their uncertainties from the marginalized
posterior distributions as the 15.9\%, 50\%, and 84.1\% percentiles. The
values of the key parameters with their uncertainties are given in Table
\ref{tab:Properties}. Note that since the errors in this case are symmetric,
we report the average between the 15.9\% and 84.1\% percentiles in the table.

Car~II is best fit by a mildly elliptical profile with $e=0.34\pm0.07$
and an elliptical half-light radius of $r_h=8\farcm7\pm0.8$, while Car~III is much
smaller at $r_h=3\farcm75\pm1$. It also appears to be significantly
elongated with $e=0.55\pm0.18$. The best-fit spatial models are shown
in Figure~\ref{fig:spatialfit}. The left panel of the Figure shows the
density map of the stars inside a mask created using the best-fit
isochrone of Car~II. We note that while this mask is not ideal for
showing Car~III, due to differences between the isochrones of Car~II and Car~III, 
the overlap between the two populations is large enough to
reveal both overdensities on the map. In the right panel, we show
stellar density contours together with the half-light ellipses of the
best-fit models. We also show the BHB candidates selected within 0.3
magnitudes from the BHB ridge-line of Car~II (III) in blue (cyan) and the RR Lyrae stars in orange. It is 
remarkable 
that BHB stars appear to be distributed more diffusely than the bulk of
the other stellar populations (such as RGB and MSTO). Note that the
spread out appearance of the BHB candidate stars, together with the
elongation of both satellites in the same direction might indicate
tidal disruption.

The final confirmation of the genuine nature of the Car~II and III system is
given in Figure~\ref{fig:hess}, where we show the CMD and Hess difference
diagrams for each object. Panels in the left column show the CMDs within
$15\arcmin$ for Car~II and $5\arcmin$ for Car~III, while the middle panels give
the CMDs of the foreground stars from a region of the same area but at least $30\arcmin$ away from the center of each satellite. Finally the
right panels present the foreground subtracted Hess difference diagrams. The
shapes of the spatial regions that define the Hess difference diagram as well
as the isochrones (shown in red lines) are dictated by the best-fit models
described above. The blue (cyan) points show BHB candidates for Car~II
(Car~III), and the red dashed line is the M92 BHB ridge-line shifted to the
distance of the isochrone. Note that the BHBs present in both satellites are
not considered in either the modeling or the search, and therefore provide
independent confirmation of Car~II and III and their distance estimations. Both
satellites have a clear MSTO - where very little background is expected, as shown by the background CMD - and
are well described by the best-fit isochrones. As evident from the CMDs and
the model isochrones, CarII and III clearly possess distinct stellar
populations. Specifically, while both stellar population are consistent with
an old ($\sim10$ Gyr) and metal poor ($[$Fe/H$]=-1.8\pm0.1$) populations,
Car~II has a distance modulus of $17.79\pm0.05$, placing it at $\sim36.2 \pm
0.6$\,kpc, but Car~III, on the other hand, has a distance modulus of
$17.22\pm0.10$, which is equivalent to a distance of $\sim 27.8 \pm 0.6$\,kpc.

Finally, we also measure the luminosity of the two newly discovered
satellites. The number of stars with $g,r<23$ from the best-fit model
is $709\pm58$ and $156\pm32$ in Car~II and Car~III respectively. Combining the best-fit isochrones with a Chabrier initial mass function, we
can estimate the absolute luminosity of each object. We measure
$M_V=-4.5\pm 0.1$ for Car~II and $M_V =-2.4\pm 0.2$ for Car~III,
confirming that Car~III is $\sim7$ times fainter than its
neighbor.

\section{Discussion and Conclusions}\label{sec:conclusions}

MagLiteS has explored uncharted regions of the sky in pursuit of an entourage of smaller satellites of the Magellanic
Clouds. Using this dataset, we have discovered two new ultra-faint satellites in the vicinity of the LMC. Named after the constellation in which they are located, Carina~II and III
are separated by only a few arcminutes, making them a tight pair on the sky. 
Analysis of their stellar populations, supported by the presence
of prominent MSTO features, reveals that they are relatively close to
each other physically (separated by $\sim 10$\,kpc), but
perhaps more interesting is that they are only $\sim 20$\,kpc from the
LMC. Whether this association, with the LMC and between the Carinas,
is physical or circumstantial is key to revealing their origin, and
could be an important ingredient in measuring the properties of the
LMC itself and the role of its satellites in the Milky Way halo.

Figure~\ref{fig:mvrh} shows Car~II and III in the context of the
structural properties of the satellite population of the Milky Way. Both
satellites lie in a region of size-luminosity space occupied by the ultra-faint
dwarfs, close to the detection limit of current surveys. It should be
noted that signs of tidal disruption, as hinted by the highly elliptical
profile of Car~III, and a diffuse population of BHB
candidates present in the case of Car~II, could lead to mildly inflated size
measurements. Even in this scenario, both Carinas appears to be much too
diffuse to be globular clusters, although for Car~III, its smaller physical
size may actually be consistent with the hypothesis of a disrupting
globular cluster. Despite the strong structural evidence for Car~II, the robust determination
as to whether Car~II and Car~III are indeed dwarf galaxies or
extended/tidally disrupting globular clusters can only be decided
through the analysis of the kinematics and/or metallicities of their stellar members \citep[see ][for a spectrocopic anaylis]{carinali}.

\subsection{Anisotropy in the DES/MagLiteS satellite distribution}

\begin{figure}
    \includegraphics[width=0.98\columnwidth]{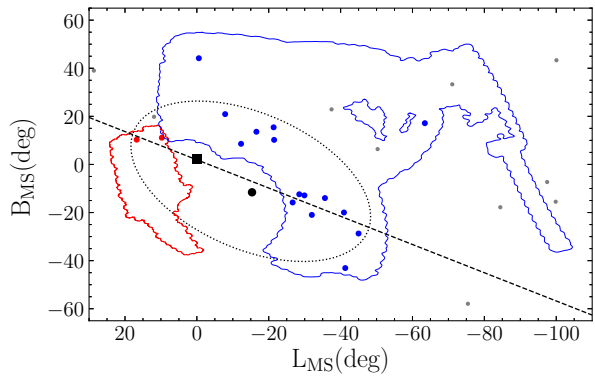}
    \caption{ On-sky positions of satellites within 300 kpc from DES (blue), MagLites
      (red, note Car~II and III are merged into a single point), the LMC (black
      square), and SMC (black circle) shown in Magellanic Stream
      co-ordinates \citep{2008ApJ...679..432N}. Other satellites in the region are shown in gray. We also show the
      footprints of DES/MagLiteS in blue/red.  The dashed line is the
      best fitting line to the positions of the MagLites satellites,
      Magellanic Clouds, and the 7 DES satellites with $B_{MS}<0$.
      The dotted ellipse represents an alternative, broader model
      distribution.}  
    \label{fig:onsky}
\end{figure}

The distribution of satellites discovered in DES is highly
anisotropic.  DES satellites preferentially lie towards the southern
edge of the survey, bordering the Magellanic Clouds, providing some
evidence for a physical association between these groups
\citep{Koposov2015,DrlicaWagner2015}.  Figure~\ref{fig:onsky} hints at
a further anisotropy in the distribution of dwarf galaxies in the
vicinity of the Magellanic Clouds.  It shows the positions of all
dwarfs and dwarf candidates within 300 kpc discovered in DES (blue circles) and
MagLiteS (red circles). Other satellites in the region are displayed as gray circles.  All three MagLiteS discoveries are well
aligned with the line connecting the LMC and SMC.  Additionally, 7
of the DES discoveries seem to fall along the same tight linear
sequence on the sky.  We add to Figure~\ref{fig:onsky} the best
fitting line to this subset of the DES satellites, the MagLiteS
satellites, and the Magellanic Clouds.  This line passes through the
position of the LMC, while the standard deviation of vertical
distance of the sample from this line is only 3 degrees.  This
elongation of the observed satellite distribution along the LMC-SMC
separation vector is an interesting curiosity. Hydra~II, which has been associated with the LMC leading arm \citep{2015ApJ...804L...5M}, sits at the opposite direction of the planar anisotropy traced by the DES and \maglites dwarfs.

What could the nature of this anisotropy be?  An on-sky linear
alignment could, of course, correspond to a 3D planar structure.
Prior to any MagLiteS discoveries, J16 considered the
group of 7 DES satellites with negative $B_{MS}$, finding that they
lie in a plane with rms thickness 2.7 kpc, which also contained both
the LMC and SMC.  Pic~II, Car~II and Car~III have distances from the
plane, as defined in that work, of 2.0 kpc \citep{Drlica-Wagner2016}, 1.8 kpc, and 1.0 kpc respectively,
i.e. all less than the originally quoted rms thickness of the plane.
This may suggest that the Carinas, Pic II and half of the DES
satellites may comprise a \emph{Magellanic satellite plane}, though
such a conclusion is highly speculative.  When \emph{all} of the DES and MagLiteS satellites are considered together
their distribution may just be described by a broad ellipse, elongated
in the direction of the line connecting LMC and SMC (e.g. the dotted
ellipse in Figure~\ref{fig:onsky}), rather than a thin plane.  For
now, we simply note this interesting feature in the observed
distribution, but defer quantitative analysis of the exact nature of
this anisotropy to a future work.

% \begin{figure}
%     \includegraphics[width=0.98\columnwidth]{img/plane_new.eps}
%     \caption{Position of satellites around the MCs in a rotated galactocentric
%     cartesian frame. Cyan/white points shows DES satellites that are on/off
%     the plane, purple shows the position of PicII, and the red points shows
%     the position of the Carina's. The sun is marked with the yellow star, the
%     plane is traced with the cyan line in the bottom left panel. The density
%     map shows the volume sampled by DES in blue and MagLiteS in red. The top
%     right histogram shows the expected RMS for groups of 7 satellites as
%     expected by the distribution of satellites in J16 model.
%     Note that the three new satellites lie closer than expected from the
%     distribution of the original satellites forming the plane. }
%     \label{fig:plane}
% \end{figure}
%

\section*{Acknowledgments} 

Support for G.T. is provided by CONICYT Chile.  We are grateful for
Director's Discretionary time on the Blanco 4m with DECam. CG
acknowledges support by the Spanish Ministry of Economy and
Competitiveness (MINECO) under grant AYA2014-56795-P. The research
leading to these results has received funding from the European
Research Council under the European Union's Seventh Framework
Programme (FP/2007-2013) / ERC Grant Agreement n. 308024. B.C.C. acknowledges the support of the Australian Research Council through Discovery project DP150100862.

Funding for the DES Projects has been provided by 
the U.S. Department of Energy, 
the U.S. National Science Foundation, 
the Ministry of Science and Education of Spain, 
the Science and Technology Facilities Council of the United Kingdom, 
the Higher Education Funding Council for England, 
the National Center for Supercomputing Applications at the University of Illinois at Urbana-Champaign, 
the Kavli Institute of Cosmological Physics at the University of Chicago, 
the Center for Cosmology and Astro-Particle Physics at the Ohio State University, 
the Mitchell Institute for Fundamental Physics and Astronomy at Texas A\&M University, 
Financiadora de Estudos e Projetos, Funda{\c c}{\~a}o Carlos Chagas Filho de Amparo {\`a} Pesquisa do Estado do Rio de Janeiro, 
Conselho Nacional de Desenvolvimento Cient{\'i}fico e Tecnol{\'o}gico and the Minist{\'e}rio da Ci{\^e}ncia, Tecnologia e Inovac{\~a}o, 
the Deutsche Forschungsgemeinschaft, 
and the Collaborating Institutions in the Dark Energy Survey. 
The Collaborating Institutions are 
Argonne National Laboratory, 
the University of California at Santa Cruz, 
the University of Cambridge, 
Centro de Investigaciones En{\'e}rgeticas, Medioambientales y Tecnol{\'o}gicas-Madrid, 
the University of Chicago, 
University College London, 
the DES-Brazil Consortium, 
the University of Edinburgh, 
the Eidgen{\"o}ssische Technische Hoch\-schule (ETH) Z{\"u}rich, 
Fermi National Accelerator Laboratory, 
the University of Illinois at Urbana-Champaign, 
the Institut de Ci{\`e}ncies de l'Espai (IEEC/CSIC), 
the Institut de F{\'i}sica d'Altes Energies, 
Lawrence Berkeley National Laboratory, 
the Ludwig-Maximilians Universit{\"a}t M{\"u}nchen and the associated Excellence Cluster Universe, 
the University of Michigan, 
{the} National Optical Astronomy Observatory, 
the University of Nottingham, 
the Ohio State University, 
the University of Pennsylvania, 
the University of Portsmouth, 
SLAC National Accelerator Laboratory, 
Stanford University, 
the University of Sussex, 
and Texas A\&M University.

This manuscript has been authored by Fermi Research Alliance, LLC under Contract No. DE-AC02-07CH11359 with the U.S. Department of Energy, Office of Science, Office of High Energy Physics. The United States Government retains and the publisher, by accepting the article for publication, acknowledges that the United States Government retains a non-exclusive, paid-up, irrevocable, world-wide license to publish or reproduce the published form of this manuscript, or allow others to do so, for United States Government purposes.

This research has made use of the APASS database, located at the AAVSO web site. Funding for APASS has been provided by the Robert Martin Ayers Sciences Fund.

BCC acknowledges the support of the Australian Research Council through Discovery project DP150100862.

GT acknowledges support from the Ministry of Science and Technology grant MOST 105-2112-M-001-028-MY3, and a Career Development Award (to YTL) from Academia Sinica.

\bibliographystyle{mn2e}
\bibliography{biblio}

\appendix
\section{Other periodic variable stars}
\label{sec-othervariables}

Table~\ref{tab-othervariables} contains information about the other 46 periodic variables detected in the field of view of Carina~II and Carina~III. The table contains ID, RA, DEC, period (in days) and amplitudes and mean magnitudes in $gri$. Notice that all but one of these stars are well outside the half-light radius of the dwarf galaxies and hence, they are expected to be non-members. Eclipsing binary star MagLiteS\_J073833.89-575638.1 is located within the $r_h$ of Car III but with no distance information is not possible to know if it is a member or just a chance alignment.

\begin{table*}
\centering
\caption{Other periodic variable stars in the FoV of Carina~II and Carina~III}
\label{tab-othervariables}
\begin{tabular}{lccccccccc} 
\hline
ID & RA     & DEC    & Period & $\Delta_g$ & $g$    & $\Delta_r$ & $r$     & $\Delta_i$ & $i$ \\ 
   & (deg) & (deg)   & (d)    & (mag)      & (mag)  & (mag)      & (mag)   & (mag)      & (mag) \\
\hline
MagLiteS\_J073209.60-575646.6 & 113.040017 & -57.94629 & 0.34764 & 0.55 & 16.61 & 0.58 & 15.81 & 0.57 & 15.50 \\ 
MagLiteS\_J073229.74-581043.7 & 113.123937 & -58.17880 & 0.31109 & 0.30 & 19.07 & 0.28 & 18.29 & 0.29 & 17.97 \\ 
MagLiteS\_J073308.82-575340.8 & 113.286750 & -57.89467 & 0.37512 & 0.17 & 15.85 & 0.13 & 15.25 & 0.12 & 15.05 \\ 
MagLiteS\_J073315.25-580605.3 & 113.313540 & -58.10146 & 0.14089 & 0.47 & 20.70 & 0.34 & 20.08 & 0.33 & 19.84 \\ 
MagLiteS\_J073338.14-574857.0 & 113.408910 & -57.81583 & 0.24591 & 0.33 & 16.05 & 0.31 & 15.58 & 0.27 & 15.43 \\ 
MagLiteS\_J073343.66-574518.9 & 113.431901 & -57.75524 & 0.32403 & 0.52 & 19.29 & 0.46 & 18.73 & 0.46 & 18.54 \\ 
MagLiteS\_J073352.46-575711.6 & 113.468576 & -57.95322 & 0.31856 & 0.17 & 15.86 & 0.15 & 15.28 & 0.12 & 15.07 \\ 
MagLiteS\_J073358.11-575602.6 & 113.492139 & -57.93406 & 0.15833 & 0.48 & 19.73 & 0.35 & 19.21 & 0.33 & 19.01 \\ 
MagLiteS\_J073411.65-580359.4 & 113.548545 & -58.06649 & 0.28094 & 0.31 & 17.28 & 0.29 & 16.49 & 0.27 & 16.20 \\ 
MagLiteS\_J073436.50-574947.2 & 113.652064 & -57.82979 & 0.41484 & 0.59 & 16.77 & 0.57 & 16.55 & 0.56 & 16.51 \\ 
MagLiteS\_J073438.78-573033.0 & 113.661576 & -57.50917 & 0.13799 & 0.29 & 19.70 & 0.28 & 19.50 & 0.28 & 19.44 \\ 
MagLiteS\_J073444.44-580138.4 & 113.685159 & -58.02732 & 0.21889 & 0.66 & 20.78 & 0.48 & 19.89 & 0.46 & 19.49 \\ 
MagLiteS\_J073445.13-572631.7 & 113.688024 & -57.44215 & 0.51143 & 0.22 & 19.92 & 0.12 & 18.97 & 0.08 & 18.53 \\ 
MagLiteS\_J073445.91-581713.8 & 113.691294 & -58.28718 & 0.35835 & 0.31 & 17.10 & 0.28 & 16.56 & 0.28 & 16.36 \\ 
MagLiteS\_J073446.35-574922.2 & 113.693109 & -57.82283 & 0.13254 & 0.44 & 20.35 & 0.19 & 19.76 & 0.18 & 19.53 \\ 
MagLiteS\_J073455.00-582253.4 & 113.729169 & -58.38151 & 0.28621 & 0.32 & 15.71 & 0.18 & 14.98 & 0.07 & 14.69 \\ 
MagLiteS\_J073456.15-582211.3 & 113.733964 & -58.36981 & 0.24194 & 0.84 & 20.48 & 0.69 & 19.67 & 0.60 & 19.31 \\ 
MagLiteS\_J073517.69-581601.2 & 113.823704 & -58.26699 & 0.09856 & 0.69 & 21.25 & 0.20 & 19.88 & 0.12 & 18.47 \\ 
MagLiteS\_J073534.40-582831.7 & 113.893341 & -58.47547 & 0.17787 & 0.17 & 16.45 & 0.20 & 15.95 & 0.25 & 15.75 \\ 
MagLiteS\_J073534.91-573536.5 & 113.895441 & -57.59348 & 0.29199 & 0.35 & 16.47 & 0.31 & 15.78 & 0.29 & 15.53 \\ 
MagLiteS\_J073545.82-581924.6 & 113.940909 & -58.32349 & 0.16011 & 0.47 & 19.97 & 0.38 & 19.50 & 0.36 & 19.31 \\ 
MagLiteS\_J073553.27-574817.0 & 113.971944 & -57.80472 & 0.36437 & 0.18 & 17.77 & 0.18 & 16.82 & 0.15 & 16.42 \\ 
MagLiteS\_J073553.50-582733.0 & 113.972928 & -58.45918 & 0.31909 & 0.61 & 16.78 & 0.56 & 15.98 & 0.60 & 15.67 \\ 
MagLiteS\_J073600.03-574943.4 & 114.000142 & -57.82873 & 0.84832 & 0.33 & 20.26 & 0.15 & 19.21 & 0.10 & 18.67 \\ 
MagLiteS\_J073602.86-574253.9 & 114.011936 & -57.71496 & 0.41908 & 0.22 & 16.43 & 0.22 & 16.24 & 0.22 & 16.19 \\ 
MagLiteS\_J073606.31-572307.8 & 114.026309 & -57.38550 & 0.16410 & 0.12 & 18.73 & 0.08 & 18.15 & 0.09 & 17.92 \\ 
MagLiteS\_J073632.63-573340.7 & 114.135944 & -57.56131 & 0.12280 & 0.12 & 18.81 & 0.13 & 18.37 & 0.11 & 18.20 \\ 
MagLiteS\_J073647.17-574048.3 & 114.196551 & -57.68009 & 0.33026 & 0.35 & 17.38 & 0.32 & 16.95 & 0.31 & 16.80 \\ 
MagLiteS\_J073648.26-573927.9 & 114.201102 & -57.65775 & 0.55454 & 0.04 & 16.88 & 0.06 & 16.52 & 0.07 & 16.38 \\ 
MagLiteS\_J073650.52-582740.4 & 114.210494 & -58.46122 & 0.52147 & 0.07 & 15.18 & 0.05 & 14.94 & 0.04 & 14.87 \\ 
MagLiteS\_J073804.20-580333.0 & 114.517499 & -58.05918 & 0.22999 & 0.64 & 18.93 & 0.55 & 18.29 & 0.52 & 18.01 \\ 
MagLiteS\_J073810.46-574918.3 & 114.543573 & -57.82174 & 0.51672 & 0.16 & 18.78 & 0.09 & 17.91 & 0.06 & 17.51 \\ 
MagLiteS\_J073821.55-582008.1 & 114.589793 & -58.33559 & 0.60013 & 0.20 & 16.27 & 0.20 & 15.96 & 0.18 & 15.85 \\ 
MagLiteS\_J073822.83-581014.3 & 114.595108 & -58.17065 & 0.16625 & 0.08 & 16.74 & 0.07 & 16.22 & 0.07 & 16.02 \\ 
MagLiteS\_J073833.89-575638.1 & 114.641226 & -57.94392 & 0.40344 & 0.30 & 17.78 & 0.29 & 17.58 & 0.28 & 17.51 \\ 
MagLiteS\_J073836.83-573145.3 & 114.653445 & -57.52924 & 0.62020 & 0.15 & 19.18 & 0.11 & 17.65 & 0.05 & 16.03 \\ 
MagLiteS\_J073839.72-581204.0 & 114.665516 & -58.20111 & 0.73806 & 0.20 & 19.30 & 0.11 & 17.97 & 0.08 & 17.19 \\ 
MagLiteS\_J073854.81-575850.1 & 114.728357 & -57.98058 & 0.18207 & 0.06 & 16.46 & 0.06 & 15.84 & 0.04 & 15.60 \\ 
MagLiteS\_J073858.32-573235.9 & 114.743007 & -57.54331 & 0.12879 & 0.55 & 20.65 & 0.47 & 20.13 & 0.49 & 19.87 \\ 
MagLiteS\_J073929.74-580919.2 & 114.873913 & -58.15533 & 0.30193 & 0.48 & 15.61 & 0.41 & 15.13 & 0.40 & 14.97 \\ 
MagLiteS\_J073938.11-575559.8 & 114.908805 & -57.93327 & 0.31649 & 0.38 & 19.94 & 0.30 & 19.89 & 0.25 & 19.90 \\ 
MagLiteS\_J073939.86-580725.0 & 114.916064 & -58.12360 & 0.11462 & 0.60 & 20.59 & 0.53 & 19.97 & 0.44 & 19.67 \\ 
MagLiteS\_J073944.42-573525.9 & 114.935070 & -57.59054 & 0.16329 & 0.55 & 18.64 & 0.52 & 18.17 & 0.49 & 17.96 \\ 
MagLiteS\_J073959.52-581942.9 & 114.997995 & -58.32859 & 0.51790 & 0.27 & 20.08 & 0.14 & 18.81 & 0.09 & 18.19 \\ 
MagLiteS\_J074018.19-575809.8 & 115.075801 & -57.96939 & 0.30735 & 0.32 & 18.81 & 0.28 & 18.18 & 0.28 & 17.92 \\ 
MagLiteS\_J074036.92-575337.5 & 115.153840 & -57.89374 & 0.27687 & 0.19 & 18.09 & 0.17 & 17.31 & 0.15 & 17.01 \\ 
\hline
\end{tabular}
\end{table*}

\label{lastpage}

\end{document}